# Deep trapping states in Cerium doped (Lu,Y,Gd)$_3$(Ga,Al)$_5$O$_{12}$ single crystal scintillators

E. Mihóková[a,*], K. Vávrů[b], K. Kamada[c], V. Babin[a], A. Yoshikawa[d], and M. Nikl[a]

[a]*Institute of Physics AS CR,v. v. i.  Cukrovarnická 10, 162 53 Prague 6, Czech Republic*
[b]*Faculty of Nuclear Sciences and Physical Engineering, Czech Technical University in Prague, Břehová 7, 115 19 Prague 1, Czech Republic*
[c]*Materials Research Laboratory, Furukawa Co. Ltd., Tsukuba 305-0856, Japan*
[d]*Institute for Materials Research (IMR), Tohoku University, Sendai 980-8577, Japan*

**Abstract**

We study deep trapping states in Ce$^{3+}$-doped garnet crystals with the composition (Lu,Y,Gd)$_3$(Ga,Al)$_5$O$_{12}$, recently shown as having remarkably high light yield. We use thermally stimulated luminescence (TSL) technique above room temperature and determine the composition Gd$_3$Ga$_3$Al$_2$O$_{12}$ as the host showing the lowest concentration of traps.  This host consistently manifest very low afterglow comparable to that of the standard BGO crystal. We also perform TSL glow peak analysis based on the initial rise technique to evaluate trap depth and other characteristics associated with TSL peaks.

[*]Corresponding author, E-mail address: mihokova@fzu.cz , Institute of Physics ASCR, Cukrovarnicka 10, 162 53 Prague 6, Czech Republic, Fax: +420 23123184, Tel.: +420 220318539

## 1. Introduction

Oxides materials based on the garnet structure are promising candidates for scintillator hosts, due to well-mastered technology originally developed for other applications, their optical transparency, and easy doping by rare-earth elements. The Ce-doped $Lu_3Al_5O_{12}$ (LuAG) single crystal was shown to be a perspective scintillating material (Nikl et al., 2000) with a relatively high density (6.7 g/cm$^3$), fast scintillation response (60-80 ns) of the $Ce^{3+}$ emission peaking around 500-550 nm, and light yield as much as 25 000 photons/MeV reported for heavily Ce-doped LuAG grown by the Bridgman method (Dujardin et al., 2010). Scintillation performance of LuAG:Ce is degraded by the presence of shallow traps (Nikl et al., 2007) whose effect can be suppressed by modifications of the band gap resulting from Ga admixture into the LuAG structure (Fasoli et al., 2011). However, too high concentration of Ga can lead to a quenching of the $Ce^{3+}$ luminescence due to a proximity of the Ce $5d_1$ excited state and the bottom of the conduction band (Ogino et al., 2009). On the other hand it is also known (Wu et al., 2007) that an admixture of large La or Gd cations into the YAG structure induces larger crystalline field that favourably down-shifts 5d states of the $Ce^{3+}$ ion within the host band gap.

Recently, remarkably high light yield was reported for tailored compositions of Ce-doped $(Lu,Gd)_3(Ga,Al)_5O_{12}$ single crystals grown by micro-pulling down technique (Kamada et al. 2011a,b). The result was achieved thanks to combination of band gap engineering and strategies for favourable $5d_1$ $Ce^{3+}$ level positioning by admixing Ga and Gd into the LuAG structure. The luminescence and scintillation mechanisms of those crystals were addressed in (Nikl et al., 2012).

In this work we study the set of Ce-doped $(Lu,Y,Gd)_3(Ga,Al)_5O_{12}$ single crystals grown by Czochralski technique. We use the thermally stimulated luminescence (TSL) technique to monitor the deep trapping states in these materials. TSL glow curves above room temperature feature several distinguished TSL peaks. We correlate TSL signal below 100 °C with the measured afterglow signal. We also use the initial rise method to determine the characteristics of traps associated with corresponding TSL peaks. In particular, we evaluate trap depths, frequency factors and room temperature lifetimes of those traps.

## 2. Samples and experimental details

The set of Ce-doped $(Lu,Y,Gd)_3(Ga,Al)_5O_{12}$ single crystals was grown by Czochralski (Cz) technique. Concentration of Ce was 1%, except for $Gd_3Ga_3Al_2O_{12}$ host where 1% and 2% Ce-doped crystals were prepared. Plates with dimensions about 5×5×1 mm were polished to an optical grade.

TSL measurements were performed by Harshaw Model 3500 Manual TLD Reader with a heating rate of 1 °C/s. Samples were irradiated at RT with the $^{60}$Co source and received the dose 2.1Gy.

Photoluminescence (PL) emission spectra in the range 270-500 K were perfod by a custom made 5000M Horiba Jobin Yvon spectrofluorometer. Excitation was realized by deuterium steady state lamp. An Oxford Instruments liquid nitrogen bath optical cryostat allowed the temperature regulation from about 270 to 500 K. Afterglow measurement was performed by the same apparatus in the spectrally unresolved mode. Excitation was realized by an X-ray tube operated at 40 kV.

## 3. Experimental results and discussion

The TSL glow curves of Ce-doped $(Lu,Y,Gd)_3(Ga,Al)_5O_{12}$ crystals after irradiation at room temperature (RT) are displayed in Fig. 1. They feature 3 to 4 distinguished peaks at about 65,115 and 170-180 °C up to about 250 °C. Above this temperature all samples feature a broad TSL structure. TSL peaks and consequently also corresponding traps seem similar in all multicomponent garnet crystals. In samples with the same concentration of $Ce^{3+}$ (1%) the highest TSL signal is observed for the garnet sample with no Gd. Substitution of Gd for the rare-earth ion (Lu or Y) lowers the TSL signal, i.e. reduces the number of traps.

Samples with Y have lower TSL compared to those with Lu. The lowest TSL signal was observed for $Gd_3Ga_3Al_2O_{12}$ (GGAG) host crystal.

To evaluate the depths of traps associated with glow peaks above room temperature we applied the *initial rise* technique (McKeever, 1985). We performed analysis of well-resolved peaks in three selected samples (see Table I). For each peak we performed two to three partial cleaning measurements with different temperatures $T_{stop}$. However, as manifested by the temperature dependence of integrated PL intensity (example for GGAG sample in the inset of Fig. 2), the $Ce^{3+}$ recombination center is quenched above RT. Consequently, before analysis, the glow curves need to be corrected for this temperature quenching of recombination center. Correction curve (solid line in the inset of Fig. 2) was obtained from the fit of PL intensity data.

The initial part of the glow peak can be approximated by an exponential function (McKeever, 1985)

$$Amp\ (T) = b + W \times \exp\ (-E/kT)\ , \qquad (1)$$

where $b$ is a constant, $E$ is the trap depth, $w$ is a preexponential factor, $k$ is the Boltzmann constant and $T$ is the absolute temperature. Numerical analysis of the data was performed by fitting the function (1) to the data obtained after partial cleaning of the glow curve and correction for the temperature quenching of recombination center.

Example of Arrhenius plot of data and fits corresponding to selected temperatures $T_{stop}$ For GGAG sample are displayed in Fig. 2 with $T_{stop}$ indicated in the figure. Corresponding trap depths resulting from fits are listed in Table I. For GAGG sample that well manifests all three analyzed TSL peaks we measured the dependence of the TSL signal on the administered radiation dose. Changing the dose by two orders of magnitude we did not observe any shift of the TSL peak positions (see Fig. 3). Therefore one may assume that the traps corresponding to TSL peaks follow the first order recombination kinetics. As a result, we calculated associated frequency factors using the formula relating the frequency factor $s$, the heating rate $\beta$ (1 K/s) and the temperature maximum of the TSL peak $T_m$ (McKeever, 1985):

$$\beta E / kT_m^2 = s \times \exp\ (-E/kT_m). \qquad (2)$$

The trap depth $E$ is taken from the initial rise evaluation, in particular, for each trap we used average values indicated in Table I.

The lifetime of the trap $\tau$ at the temperature $T$ can be calculated as (McKeever, 1985):

$$\tau = s \times \exp\ (E/kT)\ . \qquad (3)$$

The values of frequency factors $s$ and lifetimes $\tau$ at room temperature are also listed in Table I.

The traps responsible for TSL peaks around RT may affect the afterglow signal of the materials. In the current case it can be, in particular, the trap associated with the TSL peak around 64°C with a RT lifetime around 20 minutes (cf. Table I). We tried to correlate TSL signal below 100 °C to the afterglow measured at RT, displayed in Fig. 4a. Indeed, the host with the highest TSL signal, $Lu_2Y_1Ga_3Al_2O_{12}$, shows as well rather high afterglow with respect to that with the lowest TSL, namely GGAG. The latter composition not only shows an afterglow that is much lower than the former, but it is also lower than LuAG:Pr crystal and comparable to the standard $Bi_4Ge_3O_{12}$ (BGO) crystal (see Fig. 4b), known altogether as the material with considerably low afterglow (Oi et al., 1980, Moszynski et al., 1981,). The low afterglow of GGAG is also consistent with the highest light yield observed for this host (Prusa et al., 2012).

## 4. Conclusions

We performed the first study of deep trapping states of Ce-doped $(Lu,Y,Gd)_3(Ga,Al)_5O_{12}$ single crystals grown by Czochralski technique. TSL up to 300 °C features three to four resolved peaks that follow first order recombination kinetics. Characteristic parameters of corresponding traps were determined by the initial rise technique. The lowest TSL signal within the group of studied crystals is observed for the $Gd_3Ga_3Al_2O_{12}$ host. In particular, TSL below 100°C, well correlates with extremely low afterglow, comparable to the standard BGO crystal, as well as very high light yield determined by an independent study. These characteristics make Ce-doped $Gd_3Ga_3Al_2O_{12}$ crystal rather promising for scintillator applications.

## 5. Acknowledgements

This work was supported by the LH12150 project and by the Grant Agency of the Czech Technical University in Prague, grant No. SGS11/135/OHK4/2T/14.

**Table I.** Characteristic parameters of traps associated with TSL peaks in $Lu_2YGa_3Al_2O_{12}$ (LYGAG), $Lu_2GdGa_3Al_2O_{12}$ (LGGAG) and $Gd_3Ga_3Al_2O_{12}$ (GGAG). $T_m$, $T_{stop}$, E, $E_{ave}$, s and $\tau$ are the temperature of the TSL peak maximum, temperature of partial cleaning, trap depth, frequency factor and the TSL peak lifetime at RT, respectively.

| $T_m$ [°C] | $T_{stop}$ [°C] | LYGAG E [eV] | LGGAG E[eV] | GGAG E [eV] | $E_{ave}$ [eV] | s [$s^{-1}$] | $\tau$ at RT [hours] |
|---|---|---|---|---|---|---|---|
| 64 | 45 |  |  | 0.92 | 0.88±0.04 | ~$10^{12}$ | ~ 0.3 |
|  | 50 | 0.87 | 0.88 | 0.89 |  |  |  |
|  | 55 | 0.84 | 0.89 |  |  |  |  |
|  | 60 | 0.85 |  |  |  |  |  |
| 115 | 100 |  |  | 1.02 | 1.04±0.02 | ~$10^{12}$ | ~$10^2$ |
|  | 105 | 1.05 |  | 1.03 |  |  |  |
|  | 110 | 1.04 |  |  |  |  |  |
| 180 | 165 |  |  | 1.12 | 1.13±0.01 | ~$10^{11}$ | ~$10^4$ |
|  | 170 |  |  | 1.14 |  |  |  |

**Figure captions**

Fig. 1. TSL glow curves of $(Lu,Y,Gd)_3(Ga,Al)_5O_{12}$:Ce after γ-irradiation at RT.

Fig. 2. Arrhenius plot of the TSL glow curve of $Gd_3Ga_3Al_2O_{12}$:Ce (1%) after partial cleaning with several different temperatures $T_{stop}$ indicated in the figure. Empty circles are experimental data, solid lines are the fits of the function (2) to the experimental data. The inset shows normalized PL intensity as a function of temperature. Emission spectra were excited at 430 nm and integrated in the range of Ce emission 450-750 nm. Solid circles are experimental data, solid line is the fit to the data.

Fig. 3. TSL glow curves of $Gd_3Ga_3Al_2O_{12}$:Ce (1%) after various irradiation doses implied to the sample. The quantity of the dose is indicated in the legend.

Fig. 4. RT afterglow of Ce (1%)-doped $Gd_3Ga_3Al_2O_{12}$ and $Lu_2Y_1Ga_3Al_2O_{12}$ in (a) compared to the BGO standard and LuAG:Pr samples in (b). Excitation by an X-ray tube (40 kV) was cut-off at a certain instant, the luminescence signal was monitored before and after the cut-off. All the curves are normalized to their "X-ray on" luminescence intensity.

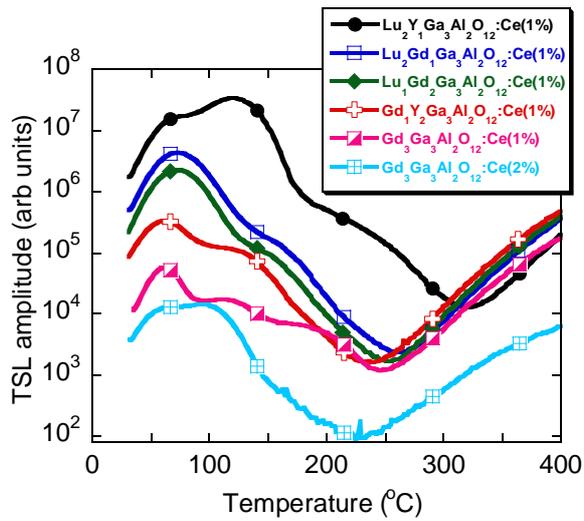

Fig. 1

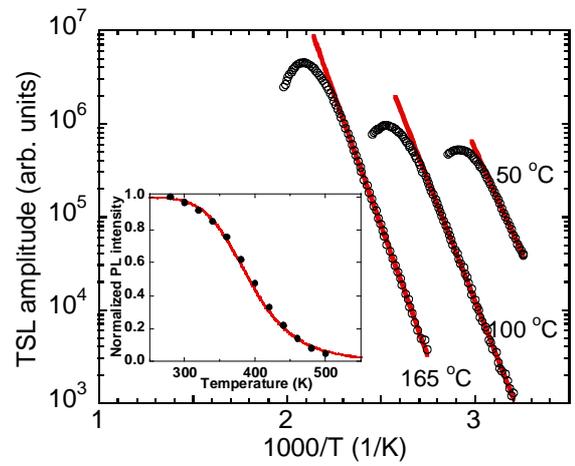

Fig. 2.

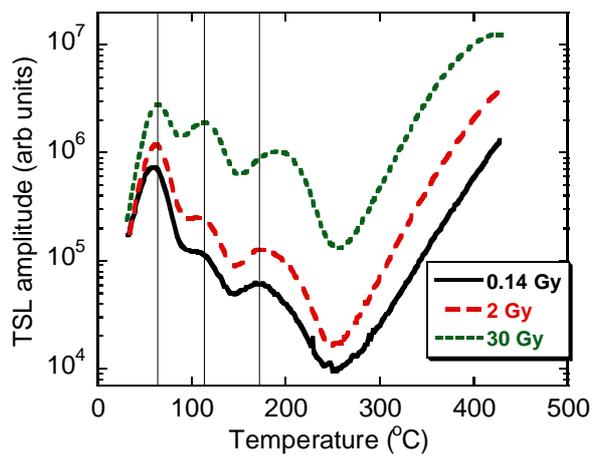

Fig. 3.

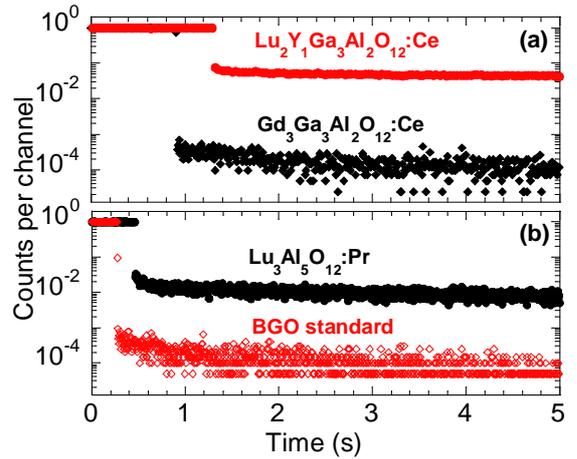

Fig. 4.